\documentclass[twocolumn,superscriptaddress,preprintnumbers,amsmath,amssymb,floatfix]{revtex4-1}
\pdfoutput=1
\usepackage{graphicx} 
\usepackage{dcolumn} 
\usepackage{bm}
\usepackage{color}
\usepackage{amsmath}
\usepackage{setspace}
\usepackage{appendix}
\usepackage[hidelinks]{hyperref}
\makeatletter
\long\def\@makecaption#1#2{%
  \par
  \begingroup
    \small
    \@parboxrestore
    \leftskip\z@
    \rightskip\z@
    \parfillskip\z@ plus 1fil
    \noindent #1. #2\par
  \endgroup
}
\makeatother
\begin{document}
\title{Machine-learning modeling of nuclear collective observables and low-lying spectra}
\author{Dan Shi}
\affiliation{School of Physical Science and Technology, Southwest University, Chongqing 400715, China}

\author{Zu-Xing Yang}
\thanks{yangzuxing@swu.edu.cn.}
\affiliation{School of Physical Science and Technology, Southwest University, Chongqing 400715, China}

\author{Xiao-Hua Fan}
\affiliation{School of Physical Science and Technology, Southwest University, Chongqing 400715, China}

\author{Zhi-Pan Li}
\thanks{zpliphy@swu.edu.cn}
\affiliation{School of Physical Science and Technology, Southwest University, Chongqing 400715, China}

\begin{abstract}

Potential energy surfaces and collective inertial functions are essential microscopic inputs for describing nuclear large-amplitude collective motions such as rotation, vibration, and fission. We develop the Nuclear Collective Generator (NCG), a machine-learning framework that predicts the collective potential and six collective inertial functions on the quadrupole deformation $(\beta,\gamma)$ plane from proton and neutron numbers and shell-effect descriptors, providing the microscopic inputs for the five-dimensional collective Hamiltonian (5DCH) used to describe low-lying spectra in even-even nuclei. The NCG combines weighted supervised learning, adversarial refinement, and ensemble averaging to improve reconstruction fidelity and prediction stability. Across 568 even-even nuclei, the reconstructed collective potentials reproduce the covariant density functional theory (CDFT) results with a mean root-mean-square deviation of 0.58~MeV. When propagated through the 5DCH solver, the NCG inputs reproduce the global systematics of equilibrium deformations, low-lying excitation spectra, and electric-quadrupole transition strengths. Near the $Z=82$ shell closure, the NCG softens the collective potential along the $\gamma$ direction, reducing the overestimated collectivity of the original CDFT+5DCH calculations and bringing the $B(E2)$ values closer to experimental data. These results demonstrate that the NCG provides an accurate surrogate for global microscopic collective calculations while retaining their essential physical content.

\vspace{1em} \noindent \textbf{Keywords:} potential energy surface; neural networks; nuclear spectrum; density functional theory

\end{abstract}

\maketitle

\section{Introduction}

The evolution of nuclear shapes is a central problem in nuclear structure physics \cite{Casten2001,Heyde2011,Meng2015}.
At the microscopic level, deformation-dependent collective potential and collective inertial functions encode the shell effects and pairing correlations that govern quadrupole collective motion \cite{Kumar1967,Ring1980,Niksic2011}. These quantities provide the microscopic inputs to the five-dimensional collective Hamiltonian (5DCH), which describes quadrupole rotational and vibrational excitations within the Bohr-Mottelson framework \cite{Bohr1970,Ring1980}.
By incorporating triaxial deformation and dynamical correlations beyond the static mean field, microscopic 5DCH calculations have successfully described low-lying spectra, electromagnetic transitions, shape coexistence, and shape phase transitions across broad regions of the nuclear chart \cite{Niksic2009,Lu2015,Delaroche2010,Yang2021,Xiang2024}.

Despite these successes, constructing the microscopic inputs of the 5DCH remains computationally demanding. In covariant density functional theory (CDFT), constrained relativistic mean-field calculations must be performed independently at many points on the $(\beta,\gamma)$ deformation mesh to determine the collective potentials and inertial functions \cite{Yang2021,Lu2015}.  Beyond the computational cost, the predictive accuracy of the current 5DCH framework is also limited by several physical approximations. The collective inertia parameters are evaluated within the cranking (Inglis--Belyaev) approximation, which neglects time-odd mean-field contributions and dynamical residual interactions, leading to systematic uncertainties in the vibrational mass parameters.
Furthermore, the deformation space is typically restricted to the quadrupole $(\beta,\gamma)$ plane, leaving out higher-order multipole degrees of freedom and pairing fluctuations that can affect both the collective potential topology and the collective dynamics.

Machine learning provides a promising route to reduce the computational burden of constructing these microscopic inputs, and offers the potential to refine the collective inertia parameters and to extend the deformation space to higher-order multipoles. Neural networks have already been applied to a variety of nuclear observables, including masses \cite{Niu2018,Ma2020}, charge radii \cite{Utama2016,Dong2022}, excited states \cite{Wang2022}, decay half-lives \cite{Saxena2021,Niu2019}, fission yields \cite{Wang2019}, and neutron skins \cite{Adhikari2021}. They have also been used to reconstruct deformation-dependent energy surfaces and collective inertial parameters for subsequent 5DCH calculations \cite{Lasseri2020}. However, predicting such collective fields is more demanding than predicting a single scalar observable: local minima, barriers, curvatures, and softness of the collective potential can strongly influence collective wave functions and spectroscopic observables. A useful surrogate model must therefore reproduce not only global trends but also the local structures relevant to collective dynamics.

In this work, we develop a machine-learning framework to reconstruct the complete set of seven deformation-dependent inputs required by the 5DCH from the proton number, neutron number, and shell-effect descriptors. These inputs consist of the collective potential, three moments of inertia, and three vibrational mass parameters. Weighted supervised learning is used as the primary training strategy, followed by adversarial refinement \cite{Goodfellow2014,Radford2015} to improve the local structures of the reconstructed fields. An ensemble of independently trained networks is further employed to enhance robustness and suppress spurious fluctuations \cite{NeurIPS2017}. The reconstructed quantities are then propagated through the 5DCH solver, and their physical reliability is assessed through low-lying excitation energies and $B(E2)$ transition strengths. The results are compared with the original CDFT+5DCH calculations and available experimental data, providing a direct test of whether the machine-learned collective potentials retain the information required for quantitative spectroscopy.

The remainder of this paper is organized as follows. Section \ref{sec2} introduces the 5DCH method and the machine-learning framework. Section \ref{sec3} validates the reconstructed collective quantities and the resulting spectroscopic observables. Section \ref{sec4} discusses the global systematics and deviations from the original CDFT+5DCH calculations. A summary and outlook are given in Sec. \ref{sec5}.

\section{Theoretical Framework \label{sec2}}

This section introduces the microscopic CDFT+5DCH framework used to generate the training targets, followed by the construction of the dataset and the machine-learning model used to reconstruct the collective quantities.

\subsection{Microscopic five-dimensional collective Hamiltonian}

Low-energy collective phenomena can be described within nuclear energy-density-functional theory. Self-consistent mean-field calculations constrained by multipole moments generate intrinsic states that break rotational symmetry \cite{Ring1980,Bender2003,Meng2006}. Such symmetry breaking captures static correlations associated with nuclear deformation, whereas a quantitative description of excitation spectra and electromagnetic transitions requires additional dynamical correlations associated with collective fluctuations and the restoration of broken symmetries. In this work, we employ the microscopic five-dimensional collective Hamiltonian (5DCH) to account for shape mixing and the restoration of rotational symmetry, with its microscopic inputs obtained from constrained CDFT calculations \cite{Niksic2009}. The particle-hole channel was described by the point-coupling functional PC-PK1 \cite{Zhao2010}, while a separable pairing force was employed in the particle-particle channel \cite{Tian2009}. Calculations over the quadrupole deformation plane $(\beta,\gamma)$ provided the total energies, quasiparticle spectra, and wave functions required to construct the collective Hamiltonian.

The constrained energy functional is written as
\begin{equation}\label{eq:constraint}
E'=E_{\rm tot}+\sum_{\mu=0,2}C_{2\mu}\left(\langle\hat Q_{2\mu}\rangle-q_{2\mu}\right)^2,
\end{equation}
where
\begin{equation}
 E_{\rm tot}=E_{\rm RMF}+E_{\rm pair}^{(p)}+E_{\rm pair}^{(n)}+E_{\rm c.m.},
\end{equation}
$E_{\rm RMF}$ and $E_{\rm pair}^{(n,p)}$ denote the self-consistent relativistic mean-field energy and the neutron and proton pairing energies, respectively. 
$E_{\rm c.m.}$ is the center-of-mass correction to the total energy. The detailed formalism can be found in Ref.~\cite{Niksic2009}.

The quadrupole operators are defined as
\begin{equation}
\begin{aligned}
\hat Q_{20} &= 2z^2-x^2-y^2, \\
\hat Q_{22} &= x^2-y^2.
\end{aligned}
\end{equation}
$q_{2\mu}$ in Eq. (\ref{eq:constraint}) is the prescribed value of the multipole moment and $C_{2\mu}$ is the corresponding stiffness constant \cite{Ring1980}. 
Scanning the $(\beta,\gamma)$ deformation plane yields the total energy surface, single-particle energies and wave functions, and occupation probabilities. The collective inertia parameters and zero-point energies are then evaluated over the full deformation space within the cranking approximation. Then we can obtain the 5DCH:
\begin{equation}\label{5DCH}
\begin{aligned}
& \hat{H}(\beta,\gamma,\Omega) \\
&= -\frac{\hbar^2}{2\sqrt{wr}} \Bigg\{\frac{1}{\beta^4} \left[\frac{\partial}{\partial\beta}\sqrt{\frac{r}{w}}\beta^4 B_{\gamma\gamma}\frac{\partial}{\partial\beta} - \frac{\partial}{\partial\beta}\sqrt{\frac{r}{w}}\beta^3 B_{\beta\gamma}\frac{\partial}{\partial\gamma}\right] \\
&\quad + \frac{1}{\beta\sin 3\gamma}\left[-\frac{\partial}{\partial\gamma}\sqrt{\frac{r}{w}}\sin 3\gamma B_{\beta\gamma}\frac{\partial}{\partial\beta} + \frac{1}{\beta}\frac{\partial}{\partial\gamma}\sqrt{\frac{r}{w}}\sin 3\gamma B_{\beta\beta}\frac{\partial}{\partial\gamma}\right]\Bigg\} \\
&\quad + \frac{1}{2}\sum_{k=1}^{3}\frac{\hat{J}_k^2}{\mathcal{I}_k} + V_{\text{coll}}(\beta,\gamma),
\end{aligned}
\end{equation}
where $V_{\text{coll}}$ is the collective potential including zero-point energy corrections, $B_{\beta\beta}$, $B_{\beta\gamma}$, and $B_{\gamma\gamma}$ are the vibrational mass parameters, $\mathcal{I}_k$ are the moments of inertia, and $w$ and $r$ are metric factors \cite{Libert1999,Prochniak2004,Niksic2009}.

The 5DCH eigenvalue problem is solved by expanding the collective wave functions in a complete basis of deformation variables and Euler angles. 
The reduced electric-quadrupole transition probabilities are then calculated as
\begin{equation}
B(E2;I_i\rightarrow I_f)=\frac{1}{2I_i+1}\left|\langle I_f||\hat M(E2)||I_i\rangle\right|^2,
\end{equation}
where $I_i$ and $I_f$ denote the initial and final angular momenta, respectively, and $\hat M(E2)$ is the electric-quadrupole operator. 
Thus, excitation spectra and electromagnetic transition strengths are obtained within a single microscopic collective framework.

The dataset used in this work is generated from microscopic CDFT+5DCH calculations for 568 even-even nuclei. 
The collective coordinate space is discretized on a regular deformation mesh with $\beta\in[0,0.7]$ and $\Delta\beta=0.1$, and with $\gamma\in[0^\circ,60^\circ]$ and $\Delta\gamma=10^\circ$. 
At each deformation point, the seven collective quantities 
\begin{equation} 
\left\{ V_{\mathrm{coll}}, \mathcal{I}_1, \mathcal{I}_2, \mathcal{I}_3, B_{\beta\beta}, B_{\beta\gamma}, B_{\gamma\gamma} \right\}
\label{eq:VIB}
\end{equation} were extracted as target variables for the NCG model.

The complete dataset is randomly divided into training, validation, and held-out test subsets in proportions of 70\%, 25\%, and 5\%, respectively. 
The test subset is excluded from all model-selection steps. 
This partition mainly probes interpolation across the sampled nuclear chart; stricter region-wise extrapolation tests are deferred to future work.


\subsection{Machine-learning framework and training strategy}
\begin{figure}[ht]
  \centering
  \includegraphics[width=1\linewidth]{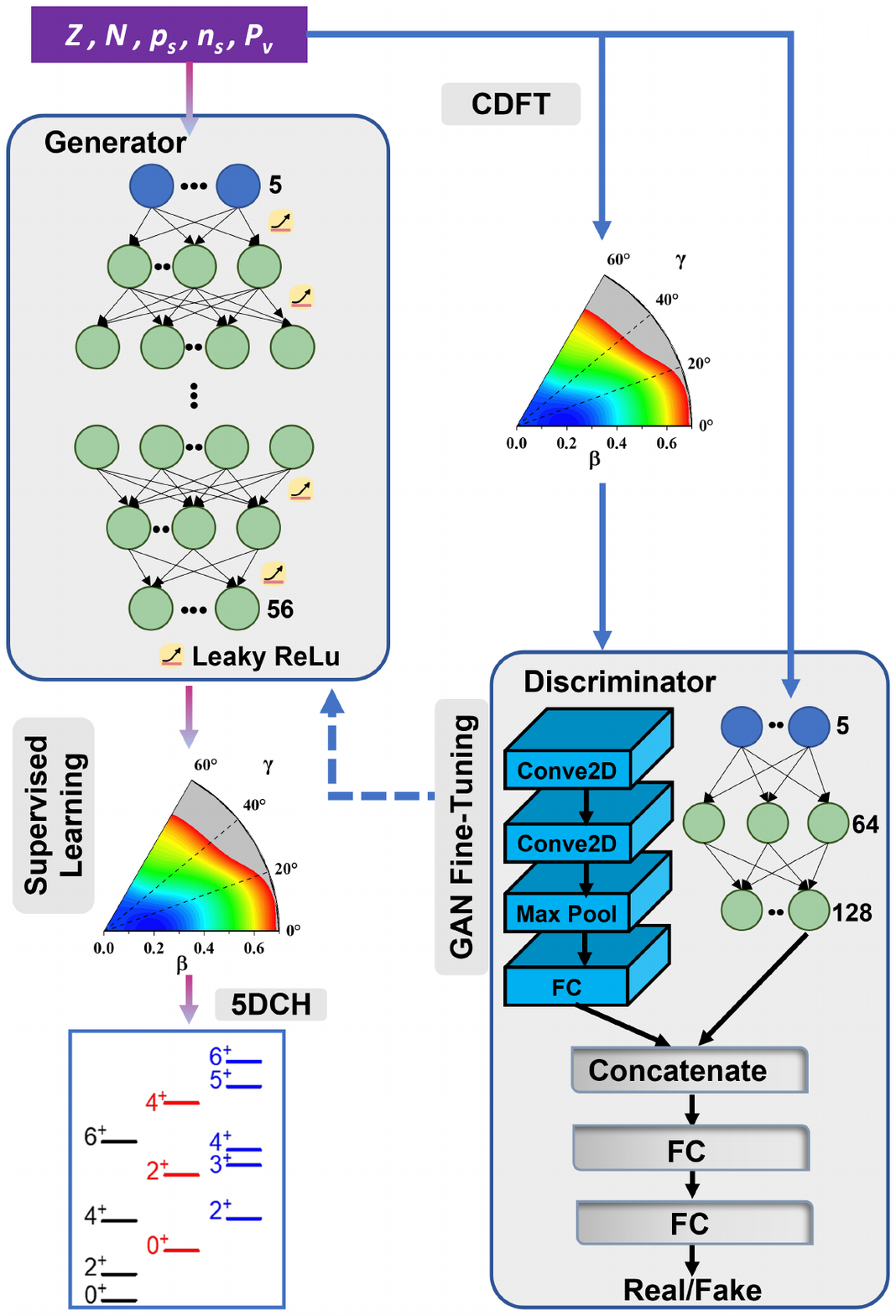}
  \caption{\label{fig1}Schematic overview of the machine-learning framework. 
  The input descriptors $(Z,N,p_s,n_s,\mathcal{P}_v)$ are mapped by generators to deformation-dependent collective fields on the $(\beta,\gamma)$ grid. 
  Supervised learning uses microscopic CDFT targets, while adversarial fine-tuning employs a conditional discriminator to regularize each predicted field. 
  The reconstructed collective quantities are then used as inputs to the 5DCH calculation of low-lying spectra.}
\end{figure}

To incorporate nuclear prior information and reduce sample complexity, the generator takes an augmented five-dimensional feature vector $\mathbf{x}= [Z, N, p_s, n_s, \mathcal{P}_v]$ as input, where $Z$ and $N$ are the proton and neutron numbers, respectively. 
Here, $p_s$ and $n_s$ denote the absolute distances to the nearest proton and neutron magic numbers ($M = 8, 20, 28, 50, 82, 126, 184$) \cite{Niu2018}, while $\mathcal{P}_v=(p_s n_s)/(p_s+n_s)$ is a combined shell-deviation descriptor that emphasizes simultaneous departures from proton and neutron shell closures. 

The overall architecture and training workflow are summarized in Fig.~\ref{fig1}. 
The figure is schematic and uses the potential-energy surface as an illustrative example. In the actual implementation, seven separate models are trained, one for each deformation-dependent quantity in Eq.~(\ref{eq:VIB}), while sharing the same architecture and optimization strategy. Each model outputs a single normalized $8\times 7$ field on the discretized $(\beta,\gamma)$ grid. Training is performed in two stages. First, the corresponding generator is pretrained with a weighted supervised loss. Second, adversarial fine-tuning may be applied to regularize the morphology of the reconstructed field and to suppress unstable local fluctuations. 
To reduce the variance associated with non-convex optimization, the final prediction for each observable and nucleus is defined as a weighted average over a committee of independently trained networks.

In the supervised-pretraining stage, the generator $G$ is optimized with a weighted mean squared error loss,
\begin{equation}
    \mathcal{L}_s=\frac{1}{N_g}\sum_{ij}
    \left(y_{\rm pre}^{ij}-y_{\rm tar}^{ij}\right)^2
    W_{ij}^2,
\end{equation}
where $N_g$ denotes the number of deformation-grid points included in the loss for a given collective quantity, 
$y_{\rm pre}=G(\mathbf{x})$ is the generator output, 
and $y_{\rm tar}$ represents the corresponding target on the $(\beta,\gamma)$ mesh.

The weighting factor is defined as
\begin{equation}
W_{ij}=y_{\mathrm{tar}}^{k}A^{l}M(i,j),
\end{equation}
where \(A=Z+N\) is the nuclear mass number and \(M(i,j)\) is a position-dependent weighting function on the discretized \((\beta,\gamma)\) grid, with \(i\) and \(j\) labeling the mesh points in \(\beta\) and \(\gamma\). 
All target quantities are normalized independently to the interval \([0,1]\) by Min-Max scaling before training, so the weighting factor remains non-negative for all observables.
In the present implementation, \(k=3\) is adopted for the collective potential, and \(k=0\) for \(B_{\beta\beta}\), \(B_{\beta\gamma}\), \(B_{\gamma\gamma}\), and the three moments of inertia. 
These values are chosen empirically based on numerical tests. 
Because the collective potential is relatively smooth and exhibits a smaller dynamic range than the other collective quantities, a larger value of \(k\) can be adopted to place greater emphasis on the weighted loss without compromising training stability. 
In contrast, the collective mass parameters and moments of inertia fluctuate much more strongly over the deformation plane. 
Setting \(k=0\) prevents these fluctuations from being overly amplified by the weighting scheme, resulting in more stable optimization and better convergence.
The mass-number exponent was tested with \(l=-1,0,\) and \(1\); \(l=-1\) yielded the lowest validation loss and was therefore adopted.
The position-dependent factor \(M(i,j)\) is implemented as a fixed observable-dependent mesh mask of order unity on the \(8\times7\) deformation grid (see the Appendix for details).
The resulting weighting scheme balances the contributions from different deformation regions and mass scales during training while preserving the overall physical trends of the target quantities.

In the adversarial fine-tuning stage, each pretrained generator is further refined within a GAN. 
Unlike the supervised pretraining stage, which minimizes the pointwise regression error, adversarial learning encourages the generated fields to reproduce the global structural characteristics of the microscopic targets. 
The generator $G$ takes the nuclear input features $\mathbf{x}$ and predicts a deformation-dependent collective quantity, while the discriminator $D$ receives both the input features and either the microscopic target field $\mathbf{y}_{\rm tar}$ or the generated field $G(\mathbf{x})$, and learns to distinguish the generated fields from the microscopic targets.
The discriminator and generator are optimized alternately by minimizing the losses
\begin{equation}
    \mathcal{L}_D = -\mathbb{E}_{(\mathbf{x},\mathbf{y}_{\text{tar}})}
    \left[
    \log D(\mathbf{x},\mathbf{y}_{\text{tar}})
    + \log \left(1-D(\mathbf{x},G(\mathbf{x}))\right)
    \right],
    \label{eq:3}
\end{equation}
\begin{equation}
    \mathcal{L}_G =
    -\mathbb{E}_{\mathbf{x}}
    \left[
    \log D(\mathbf{x},G(\mathbf{x}))
    \right].
    \label{eq:4}
\end{equation}
where \(\mathbb{E}[\cdot]\) denotes the expectation over the corresponding training samples. 
The discriminator output \(D(\mathbf{x},\mathbf{y})\in(0,1)\) represents the estimated probability that the input field \(\mathbf{y}\), conditioned on the nuclear feature vector \(\mathbf{x}\), is drawn from the microscopic training data rather than generated by the network. 
During training, \(D(\mathbf{x},\mathbf{y}_{\rm tar})\) is expected to approach 1 for real samples, whereas \(D(\mathbf{x},G(\mathbf{x}))\) approaches 0 when the generated field can still be distinguished from the microscopic target.
The first term in Eq.~(\ref{eq:3}) maximizes the discriminator response to the real samples, whereas the second term suppresses its response to generated samples. 
Equation~(\ref{eq:4}) fine-tunes the pretrained generator by increasing
\(D(\mathbf{x},G(\mathbf{x}))\), thereby encouraging the generated fields to fool the discriminator and become indistinguishable from the microscopic targets.


For each collective input, the final prediction is obtained from a weighted ensemble of \(\mathcal{N}=60\) independently trained models,
\begin{equation}
    \bar{Y}_{\mathrm{pre}}
    =
    \sum_{m=1}^{\mathcal{N}} w_m Y_{\mathrm{pre}}^{(m)},
    \qquad
    \sum_{m=1}^{\mathcal{N}} w_m = 1,
\end{equation}
where the weights are determined from the benchmark reconstruction errors,
\begin{equation}
    w_m
    =
    \frac{\epsilon_m^{-3/2}}
         {\sum_{n=1}^{\mathcal{N}}\epsilon_n^{-3/2}},
\end{equation}
with \(\epsilon_m\) denoting the mean squared error of the \(m\)-th model. 
This weighting assigns larger contributions to models with better accuracy while preserving the normalization of the ensemble.
Details of the network architecture, learning-rate settings, discriminator design, and training schedule are given in Appendix~A.

\section{Validation on representative nuclei \label{sec3}}
This section evaluates the predicted collective observables at two levels. 
The in-sample performance is first examined for $^{76}$Kr, a representative nucleus from the training and validation domain. 
Generalization is then tested on nuclei that were completely excluded from model optimization, with emphasis on whether the predicted collective potentials and inertial parameters generate physically reasonable 5DCH spectra.

\subsection{\texorpdfstring{In-sample validation for $^{76}\text{Kr}$}{In-sample validation for 76Kr}}

 \begin{figure}[ht]
  \centering
  \includegraphics[width=1\linewidth]{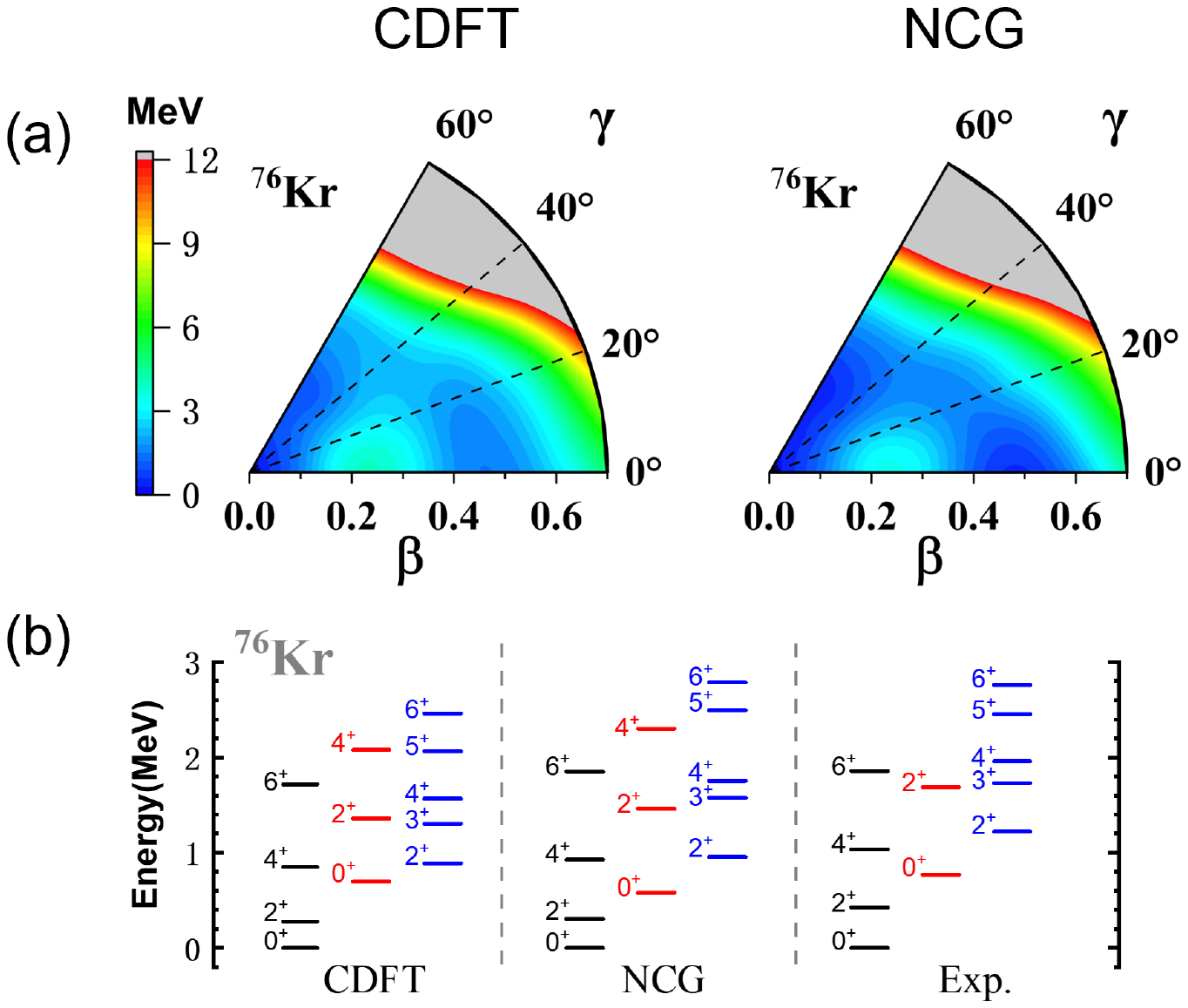}
  \caption{\label{fig2} Collective potentials and low-lying spectra of $^{76}\mathrm{Kr}$. (a) Collective potentials in the $(\beta,\gamma)$ plane obtained from CDFT (left) and NCG (right). The energy of each surface is normalized to its respective minimum, and the contour interval is 0.3 MeV. (b) Low-lying spectra calculated by 5DCH based on CDFT (left) and NCG (middle), compared with experimental data (right)~\cite{NNDC2026}. The ground-state, quasi-$\beta$, and quasi-$\gamma$ bands are shown in black, red, and blue, respectively.
} 
\end{figure}

To assess the in-sample reproducibility of the proposed model, the shape-coexisting nucleus $^{76}\text{Kr}$ is selected as a representative case from the training and validation domain (see Fig.~\ref{fig2}). This nucleus has relatively complete experimental information for the low-lying states, enabling a direct comparison among the microscopic CDFT+5DCH calculation, the NCG prediction, and experiment. Figure~\ref{fig2}(a) shows that the NCG collective potential is consistent with that from CDFT: both reveal that $^{76}$Kr exhibits a clear coexistence of a soft oblate shape and a well-deformed prolate minimum, connected through the triaxial $\gamma$ degree of freedom with a barrier of about 1~MeV. This provides reliable inputs for the 5DCH, and the resulting excitation spectra are shown in Fig.~\ref{fig2}(b). The $2^+$, $4^+$, and $6^+$ members of the ground-state band, the $\gamma$-vibrational bandhead and splittings, and the deformation-sensitive $0_2^+$ band are all reproduced. 

This in-sample comparison indicates that the model does not merely reproduce isolated values for collective potential; rather, the reconstructed seven collective inputs remain sufficiently consistent to yield a physically reasonable 5DCH spectrum that stays close to the microscopic benchmark. In addition, the reconstruction error for $^{76}\mathrm{Kr}$ is compared with the overall error distribution of the corresponding dataset and is found to lie close to the median, indicating that this nucleus represents a typical case of the model performance.

\subsection{Out-of-sample validation}

\begin{figure*}[ht]
    \centering
    \includegraphics[width=\textwidth,height=0.9\textheight,keepaspectratio]{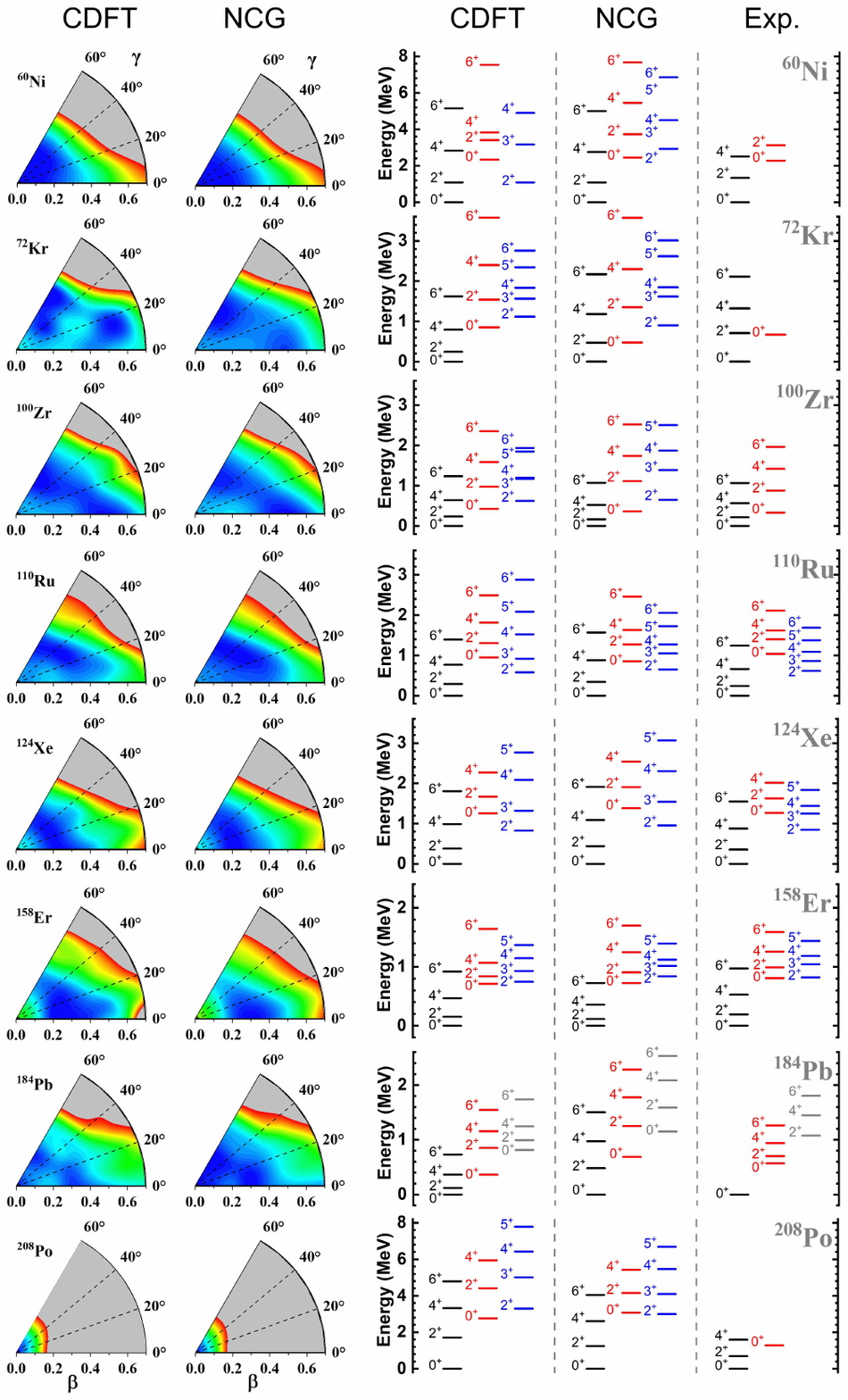}  
\caption{
Out-of-sample validation for eight held-out nuclei spanning spherical, transitional, triaxial, and shape-coexisting regions. Left: comparison of CDFT and NCG collective potentials in the $(\beta,\gamma)$ plane. Right: corresponding low-lying spectra calculated with CDFT+5DCH and NCG+5DCH, together with experimental data~\cite{NNDC2026}.
}
\label{fig:3}
\end{figure*}

Out-of-sample performance is then assessed on the held-out test subset. Eight nuclei, $^{60}$Ni, $^{72}$Kr, $^{100}$Zr, $^{110}$Ru, $^{124}$Xe, $^{158}$Er, $^{184}$Pb, and $^{208}$Po, are selected for detailed illustration. They cover different mass regions and representative deformation regimes, including spherical, transitional, triaxial, and shape-coexisting systems. The left panels of Fig.~\ref{fig:3} compare the CDFT and NCG collective potentials for these nuclei. The nearly spherical structures of $^{60}$Ni and $^{208}$Po near shell closures are preserved, and the transitional nuclei $^{124}$Xe and $^{158}$Er, whose collective potentials are very soft along the $\gamma$ and $\beta$ directions, respectively, are also well reproduced. The near-triaxial nucleus $^{110}$Ru is also well described by both CDFT and NCG. For the shape-coexisting nuclei $^{72}$Kr, $^{100}$Zr, and $^{184}$Pb, the model captures the coexistence of multiple minima and reproduces their low-lying spectra reasonably well. The NCG therefore reproduces the dominant deformation systematics of the microscopic calculations while retaining local structures relevant to collective dynamics.

The generator quality is further quantified by the root-mean-square deviation between the NCG and microscopic collective potentials. For all the nuclei in our dataset, the mean RMS deviation is 0.58 MeV. It is found that larger deviations occur mainly near shell closures and in transitional regions, where the collective potential topology changes rapidly. Although these values provide a global measure of the surface reconstruction, they do not fully determine the spectroscopic accuracy. Low-lying 5DCH spectra are particularly sensitive to the topology of the low-energy minima and to the accompanying collective inertial functions.

To examine this point, the generated collective potentials and inertial functions are used together as inputs to the 5DCH. As shown in the right panels of Fig.~\ref{fig:3}, NCG+5DCH generally preserves the agreement with experiment achieved by the original CDFT+5DCH calculation. Encouragingly, in cases where the microscopic calculation shows visible deviations, the reconstructed inputs also yield moderate improvements. For $^{72}$Kr and $^{184}$Pb, for example, the ground-state bands become less compressed and exhibit a medium-deformed or near-spherical structure. The $\gamma$-vibrational bandheads and level splittings of $^{110}$Ru and $^{158}$Er are also reasonably reproduced. These results demonstrate that the NCG collective quantities provide physically reliable inputs for 5DCH calculations of unseen nuclei.


\section{Global systematics and physical interpretation \label{sec4}}

The representative nuclei discussed above demonstrate the local accuracy of the reconstructed collective quantities. We now extend the analysis to all even-even nuclei to examine whether the NCG preserves the global systematics across the nuclear chart. The discussion begins with the predicted equilibrium deformations, followed by comparisons of excitation energies and $B(E2)$ values of low-lying collective states, and concludes with the $B(E2)$ systematics near the $Z=82$ shell closure, where the systematic refinement introduced by the NCG is most evident.

\subsection{Deformation minima across the nuclear chart}

\begin{figure}[htbp]
  \centering
  \includegraphics[width=1\linewidth]{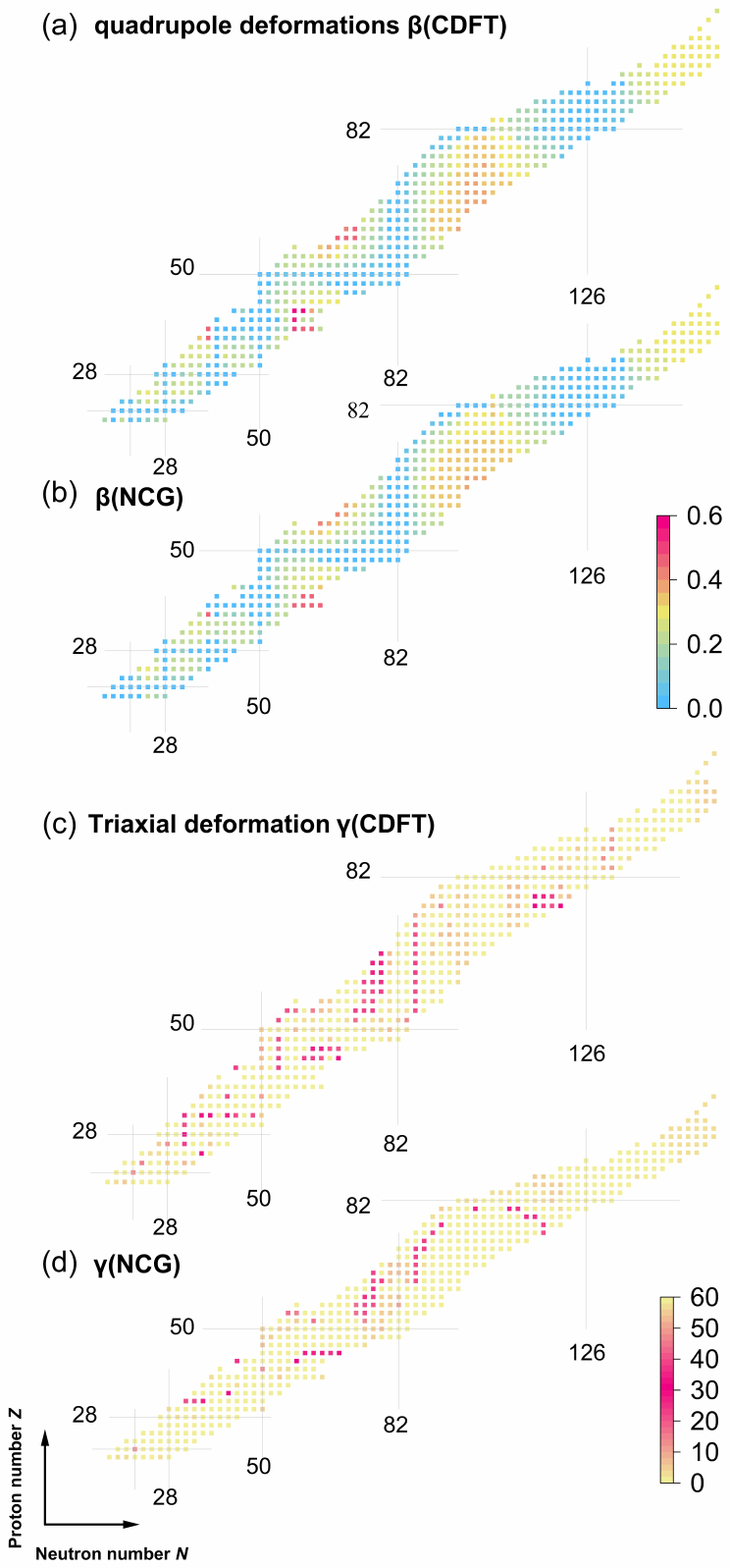}
   \caption{\label{fig4}Comparison of nuclear deformation parameters between CDFT targets and NCG predictions across the $N$-$Z$ plane. 
   Panels (a) and (b) display the quadrupole deformation $\beta$, while panels (c) and (d) show the triaxial deformation $\gamma$ at the PES minima.}
\end{figure}

Figure~\ref{fig4} compares the equilibrium deformation parameters $\beta$ and $\gamma$ obtained from the CDFT and NCG collective potentials across the $N$-$Z$ plane. Here, $\beta$ measures the magnitude of the quadrupole deformation, while $\gamma$ characterizes its triaxiality \cite{Bohr1998}. The NCG reproduces the global evolution of both quantities without introducing evident localized artifacts. Nearly spherical nuclei with $\beta\approx0$ are concentrated around the major shell closures at $Z,N=20$, 28, 50, 82, and 126. Moving toward mid-shell regions, particularly in the rare-earth and actinide sectors, the equilibrium shapes evolve toward pronounced prolate deformation with $\beta>0.3$. This spherical-to-deformed transition is well reproduced by the NCG, indicating that the shell-related input features capture the dominant evolution of collectivity across the nuclear chart. 

The $\gamma$ distributions also reproduce the main regions of nonaxial deformation, including those around $(Z,N)\approx(44,66)$, $(60,76)$, and $(76,114)$, with comparable locations and spatial extents \cite{Delaroche2010}. The residual differences in $\gamma$ arise because the collective potential of triaxially deformed nuclei is often soft along the $\gamma$ direction; consequently, the position of the absolute minimum can shift without indicating a substantial change in the underlying collective landscape \cite{Li2010}. As a smooth global approximator stabilized by ensemble averaging, the NCG tends to suppress such weak and numerically unstable local variations. 


\subsection{Excitation energies of low-lying collective states}
 \begin{figure}[htbp]
  \centering
  \includegraphics[width=1\linewidth]{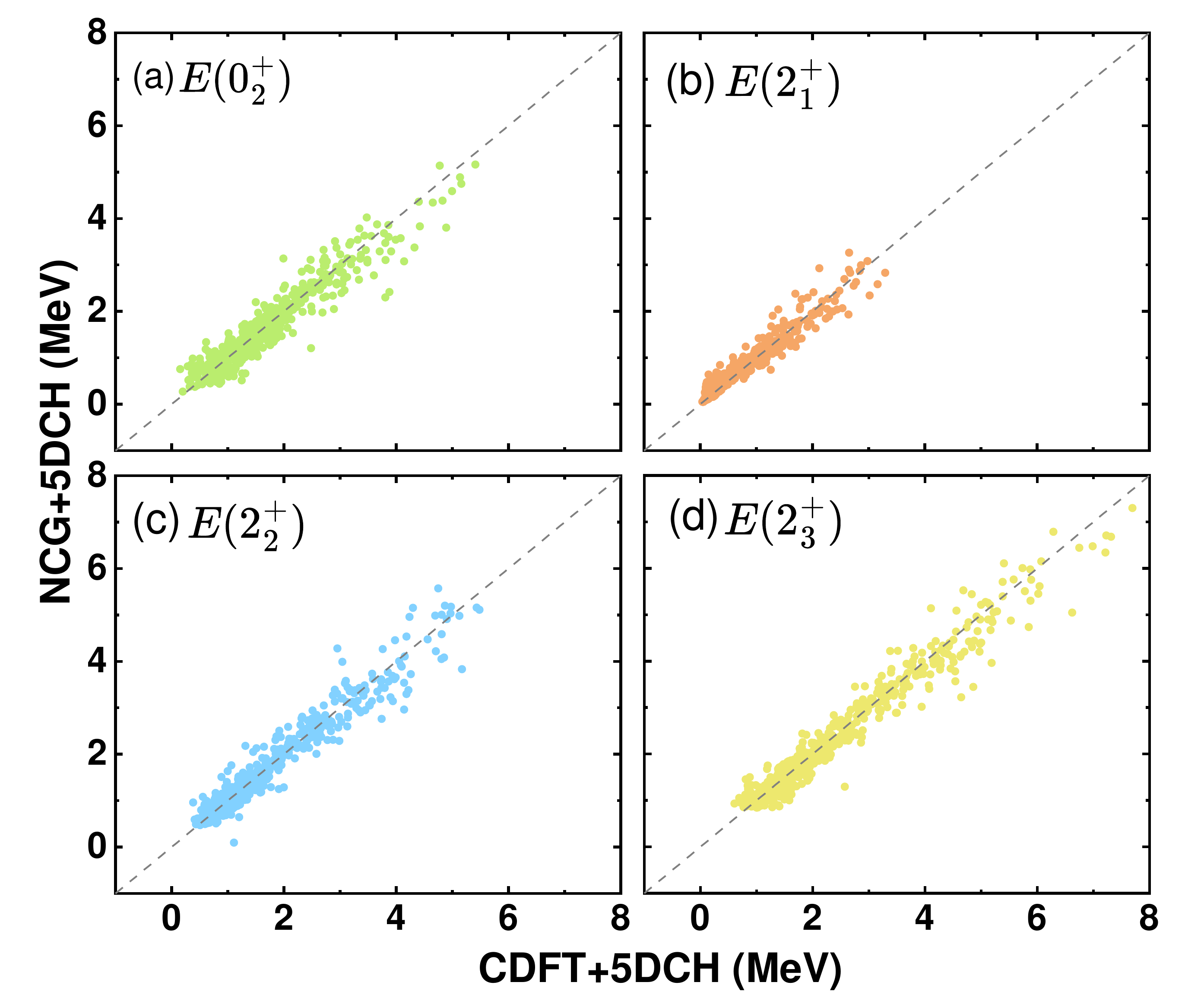}
  \caption{\label{fig5}Systematic comparisons of low-lying excitation energies across even-even nuclei. The horizontal axis shows CDFT+5DCH values, and the vertical axis shows NCG+5DCH predictions. Panels display (a) $E(0_2^+)$, (b) $E(2_1^+)$, (c) $E(2_2^+)$, and (d) $E(2_3^+)$. The gray line denotes $y=x$.}
\end{figure}

The global quality of the predicted dynamical observables is further assessed through scatter plots of $E(0_2^+)$, $E(2_1^+)$, $E(2_2^+)$, and $E(2_3^+)$. 
The $2_1^+$ energy primarily reflects the overall quadrupole collectivity and moment of inertia, whereas the $0_2^+$ and higher $2^+$ states are more sensitive to shape coexistence, $\beta$ vibration, and $\gamma$ softness. 
Most points cluster around the $y=x$ diagonal, indicating stable performance across weakly deformed, strongly deformed, and shape-coexisting nuclei.
In constructing this comparison, a small number of well-deformed light nuclei with exceptionally large deviations have been excluded from the main statistical analysis. 
These cases are likely affected by the enhanced sensitivity of light strongly deformed systems to local shell effects, pairing correlations, and collective inertial functions, and may require a dedicated treatment beyond the present global surrogate framework.

As shown in Fig.~\ref{fig5}, the correlations are tightest in the low-energy window below approximately 3~MeV, where the data density is highest. 
At higher excitation energies and near closed shells, the dispersion increases but remains bounded, with no evident divergent outliers. 
This pattern is physically reasonable, since closed-shell nuclei often have stiffer collective potentials and rapidly varying collective masses around the spherical minimum. 
Together with the analysis of the equilibrium deformation minima in Fig.~\ref{fig4}, these results show that the machine-learning framework preserves the dominant collective potential topologies and the associated 5DCH spectral systematics across the nuclear chart.

\subsection{Systematics of $B(E2)$ transition strengths}
 \begin{figure}[htbp]
  \centering
  \includegraphics[width=1\linewidth]{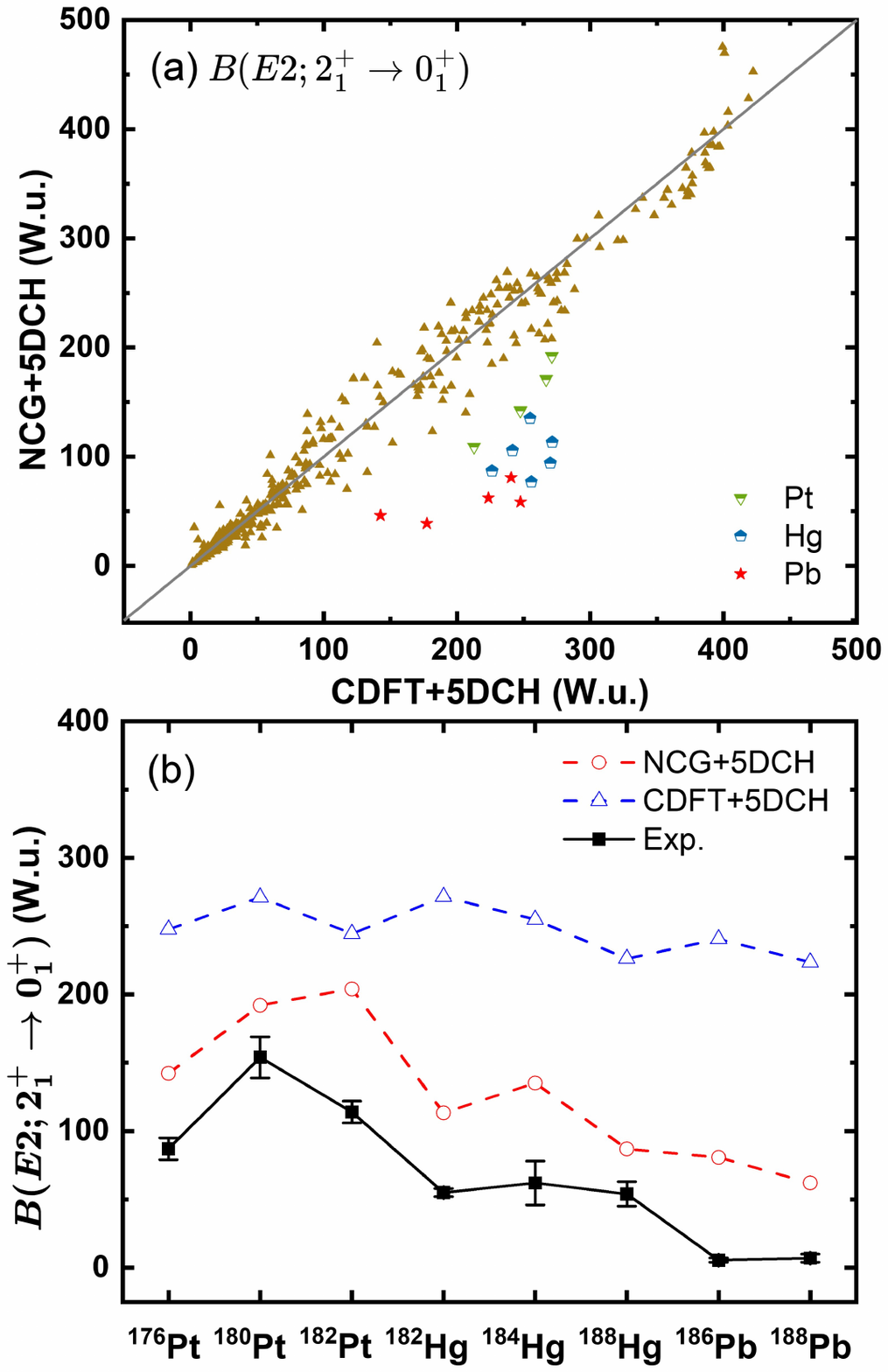}
  \caption{\label{fig6}$B(E2;2_1^+\rightarrow0_1^+)$ transition strengths. (a) NCG+5DCH versus CDFT+5DCH values across the nuclear chart, with Pt, Hg, and Pb isotopes highlighted. (b) CDFT+5DCH and NCG+5DCH results compared with experimental data~\cite{NNDC2026} along selected isotopic chains in the neutron-deficient sub-lead region.}
\end{figure}

Electromagnetic transition strengths provide an additional test of the predicted collective wave functions. Figure~\ref{fig6}(a) compares the NCG+5DCH and CDFT+5DCH values of $B(E2;2_1^+\rightarrow0_1^+)$ across the nuclear chart. Most nuclei lie close to the diagonal, demonstrating that the NCG preserves the global transition-strength systematics of the microscopic calculations. The large deviations are concentrated in the highlighted Pt, Hg, and Pb isotopes near the $Z=82$ proton shell closure, where collective observables are particularly sensitive to changes in the collective potentials and inertial functions.

A closer inspection reveals a clear and unexpected improvement in this region. As shown in Fig.~\ref{fig6}(b), the original CDFT+5DCH calculations substantially overestimate the measured $B(E2)$ values, in some cases by several factors, whereas the NCG+5DCH results markedly reduce this discrepancy while preserving the overall isotope-dependent trend. The comparison therefore demonstrates that the NCG corrects part of the systematic overestimation of quadrupole collectivity near the $Z=82$ shell closure \cite{Quan2017,Yang2023PRC}.

\begin{figure}[htbp]
  \centering
  \includegraphics[width=0.48\textwidth,height=0.9\textheight,keepaspectratio]{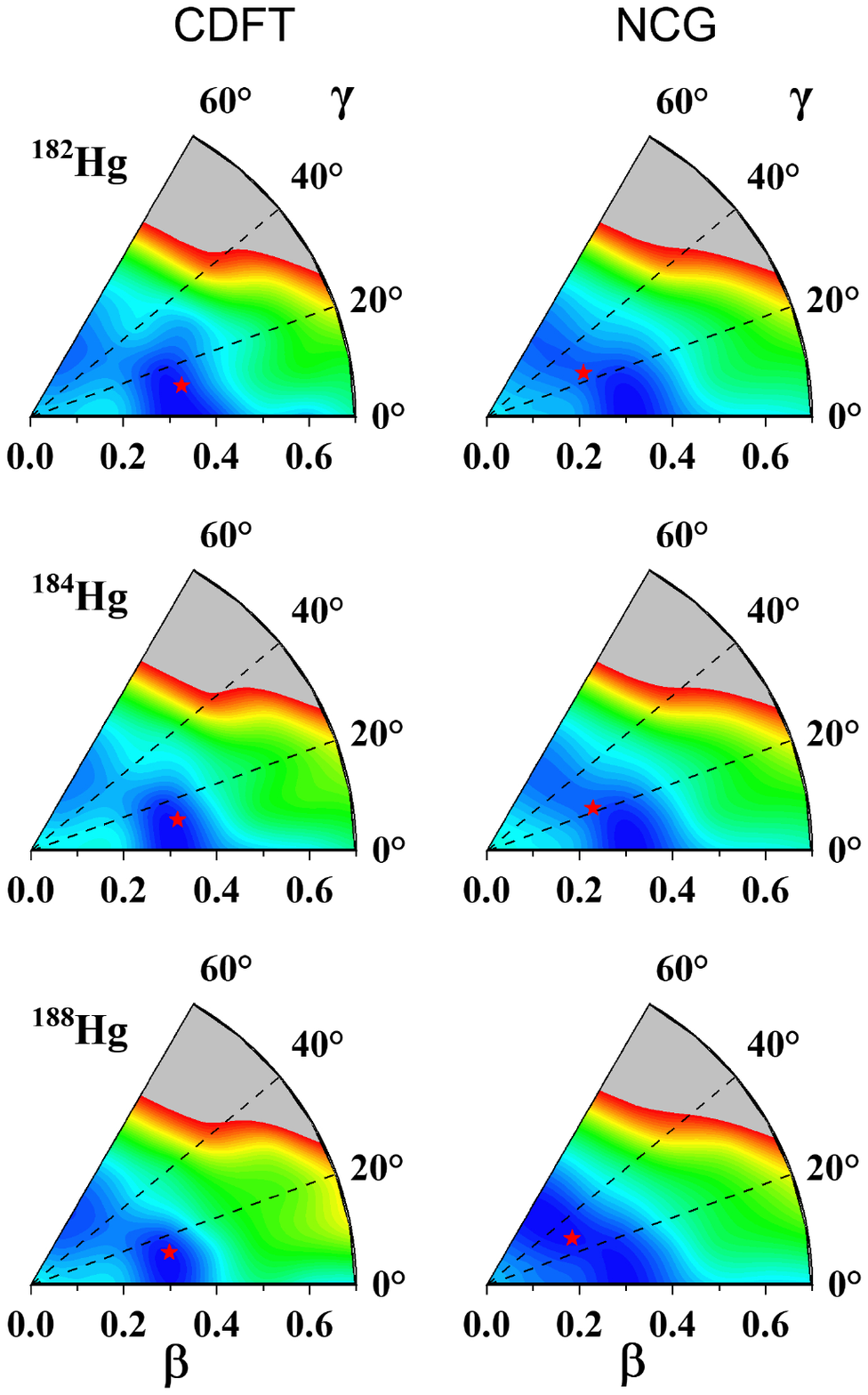}
  \caption{\label{fig7} Collective potentials in the $(\beta,\gamma)$ plane along the Hg isotopic chain, obtained from CDFT and NCG. Each surface is normalized to its own minimum, and the contour interval is 0.3~MeV. The red stars denote the $(\langle\beta\rangle,\langle\gamma\rangle)$ values of the ground $0^+_1$ states.}
\end{figure}

To understand how the NCG improves the description of $B(E2)$ in this region, we show in Fig.~\ref{fig7} the evolution of the collective potentials along the Hg isotopic chain. Compared with the original CDFT results, the NCG potentials become visibly softer along the $\gamma$ direction, which drives the ground states toward larger $\gamma$ and smaller $\beta$ deformations and thereby reduces the corresponding $B(E2)$ values. Quantitatively, the $(\langle\beta\rangle,\langle\gamma\rangle)$ values calculated by CDFT+5DCH for $^{182}$Hg, $^{184}$Hg, and $^{188}$Hg are (0.33, 11.8$^\circ$), (0.32, 12.0$^\circ$), and (0.31, 13.2$^\circ$), respectively, while they change to (0.23, 24.6$^\circ$), (0.25, 21.9$^\circ$), and (0.21, 28.8$^\circ$) in the NCG+5DCH calculations. This analysis demonstrates that the NCG better captures the evolution of the collective properties and the underlying microscopic structure in this region, providing a useful reference for the further optimization of the CDFT functional. 

\section{Summary and outlook \label{sec5}}

We have developed the NCG, a machine-learning framework for predicting the deformation-dependent collective potential, three moments of inertia, and three vibrational mass parameters on the $(\beta,\gamma)$ deformation plane. The model combines shell-related nuclear descriptors, weighted supervised learning, adversarial refinement, and ensemble averaging. Feeding these collective parameters into the 5DCH yields the experimentally observable collective excitation spectra. The predicted quantities preserve the principal systematics of the underlying CDFT calculations, including equilibrium deformations, low-lying excitation spectra, and electric-quadrupole transition strengths across spherical, transitional, well-deformed, and shape-coexisting regions.

A notable result is obtained for the Pt--Hg--Pb isotopes near the $Z=82$ proton shell closure. Here the NCG softens the collective potential along the $\gamma$ direction, driving the ground states toward larger $\gamma$ and smaller $\beta$ deformations. This reduces the systematic overestimation of collectivity in the original CDFT+5DCH calculations and brings the $B(E2)$ values closer to experiment. These results demonstrate that the NCG provides a practical surrogate for large-scale microscopic collective calculations while retaining their essential physical content.

Looking forward, the present framework, which focuses on the quadrupole $(\beta,\gamma)$ degree of freedom, can be extended in two directions. First, the deformation space can be enlarged to include the octupole deformation degree of freedom, enabling a systematic description of both positive- and negative-parity low-lying spectra~\cite{Xiang2026}. Second, the framework can be extended to describe the fission potential-energy surfaces of heavy and superheavy nuclei, with the aim of refining the fission barriers and yield distributions~\cite{Schunck2016}.

\begin{acknowledgments}
This work was supported partly by the National Natural Science Foundation of China (Grant No.12505145, No.12375126) and the Fundamental Research Funds for the Central Universities (Grant No.~SWU-KR25034).
\end{acknowledgments}

\appendix
\section{Network architecture and training details}

Seven independent generators are trained for the seven collective quantities in Eq.~(\ref{eq:VIB}), using the same architecture and optimization protocol. The main implementation settings are summarized below.

\begin{itemize}

\item \textbf{Generator architecture.}
The five-dimensional input vector
$\mathbf{x}=(Z,N,p_s,n_s,\mathcal{P}_v)$
is processed by seven fully connected hidden layers with widths
\[
32,\ 128,\ 256,\ 512,\ 1024,\ 512,\ 256.
\]
Each hidden layer uses a LeakyReLU activation with a negative slope of 0.01. The output layer contains $56$ neurons with a Sigmoid activation and is reshaped into an $8\times7$ field on the $(\beta,\gamma)$ grid. All target quantities are independently normalized to $[0,1]$ using Min--Max scaling. No explicit physical constraints are imposed on the network outputs.

\item \textbf{Weighted loss and mesh mask.}
In the supervised-pretraining stage, a fixed position-dependent mesh mask $M(i,j)$ is used in the weighted mean squared error loss, where $i$ and $j$ label the $\beta$ and $\gamma$ mesh points, respectively. 
In the present calculation, the mask is chosen as
\begin{equation}
M(i,j)=
\begin{pmatrix}
1 & 1 & 1 & 1 & 1 & 1 & 1 \\
1 & 1 & 1 & 1 & 1 & 1 & 1 \\
1 & 1 & 1 & 1 & 1 & 1 & 1 \\
1 & 1 & 1 & 1 & 1 & 1 & 1 \\
1 & 1 & 1 & 1 & 1 & 1 & 1 \\
1 & 1 & 1 & 1 & 0.5 & 0.5 & 0.5 \\
1 & 1 & 1 & 1 & 0.5 & 0.5 & 0.5 \\
1 & 1 & 1 & 1 & 0.5 & 0.5 & 0.5
\end{pmatrix}.
\end{equation}
This mask is used only as a numerical reweighting factor for the supervised loss.
\item \textbf{Supervised pretraining.}
The validation loss is monitored during training. If no improvement is observed for 200 consecutive epochs, the learning rate is reduced by a factor of 0.1 and the model is restored to the best checkpoint. Supervised training terminates when the learning rate falls below $10^{-8}$.

\item \textbf{Discriminator architecture.}
The conditional input $\mathbf{x}$ is mapped through fully connected layers
$5\rightarrow64\rightarrow128$.
The corresponding $8\times7$ target or generated field is processed by two convolutional layers with channel dimensions
$1\rightarrow16\rightarrow64$,
followed by $2\times2$ max pooling and a fully connected layer of dimension 512. The two pathways are concatenated and passed through fully connected layers
\[
640\rightarrow256\rightarrow64\rightarrow32\rightarrow1,
\]
with LeakyReLU activations and a final Sigmoid output.

\item \textbf{Adversarial fine-tuning.}
The discriminator is optimized with Adam using an initial learning rate
$\eta_D=10^{-4}$ until its total classification accuracy reaches 0.90. The generator is then optimized with
$\eta_G=10^{-7}$ while the discriminator accuracy on generated samples exceeds 0.52. The maximum number of inner iterations is 800. When this limit is reached, $\eta_D$ and $\eta_G$ are reduced by factors of 0.01 and 0.1, respectively. Only checkpoints whose validation-set mean squared error improves upon the best supervised result are retained. Training terminates when both learning rates fall below $10^{-9}$.

\item \textbf{Ensemble prediction.}
For each collective quantity, the final result is obtained from an ensemble of $N=60$ independently trained models with different random initializations and data-shuffling sequences. The ensemble output is finally transformed back to the original physical scale.

\end{itemize}

\bibliographystyle{apsrev4-1}
\bibliography{ref}
\end{document}